\documentclass[a4paper]{jpconf}
\usepackage{graphicx}
\begin{document}
\title{Recent results of the ANTARES neutrino telescope}

\author{Juan Jos\'e Hern\'andez-Rey (for the ANTARES Collaboration) }

\address{Instituto de F\'{\i}sica Corpuscular, Universitat de
  Val\`encia--CSIC, E-46100 Valencia, Spain}

\ead{Juan.J.Hernandez@ific.uv.es}

\begin{abstract}
  ANTARES, the first neutrino telescope deployed under the sea and the
  largest in the Northern Hemisphere, started to take data in
  2007. Its main goal is the detection of high energy neutrinos coming
  from cosmic sources, but other studies, such as the search for
  WIMPs, monopoles or nuclearites, are also possible.  We present here
  some of the latest results from the ANTARES neutrino telescope.
\end{abstract}

\section{The ANTARES Neutrino Telescope}

Several astrophysical objects both Galactic and extra-galactic have
been proposed as sites of acceleration of protons and nuclei, but no
conclusive experimental evidence has been obtained yet.  The decays of
mesons produced by the interactions of protons and nuclei with matter
or radiation would yield neutrinos, thus indicating the presence of
hadronic acceleration. The detection of high energy cosmic neutrinos
can therefore shed light on the origin of cosmic rays.  Several
neutrino telescopes based on the detection of muons through Cherenkov
light are at present operating worldwide.

The ANTARES Collaboration completed the construction of a neutrino
telescope in the Mediterranean Sea in May 2008, although a partial
version of the device was operating since 2007.  The telescope,
located 40 km off the Southern coast of France (42$^{\circ}$48'N,
6$^{\circ}$10'E) at a depth of 2475 m, consists in a three-dimensional
array of photomultipliers housed in glass spheres, called optical
modules, distributed along twelve lines anchored to the sea bottom and
kept taut by a buoy at the top. Each line is composed of 25 storeys of
triplets of optical modules, each housing one 10-inch photomultiplier.
The lines are subject to the sea currents and can change shape and
orientation. A positioning system based on hydrophones, compasses and
tiltmeters is used to monitor the detector geometry with an accuracy
of about $10$~cm. More details of the ANTARES telescope can be found
in~\cite{Antares}.

Muons induce the emission of Cherenkov light in sea water. The arrival
time and intensity of the Cherenkov light on the OMs are digitized
into hits and transmitted to shore.  Events containing muons are
selected from the continuous deep sea optical backgrounds due to
natural radioactivity and bioluminescence. The arrival time of the
Cherenkov photons can be determined at the nanosecond
level~\cite{TimeCalib}, allowing the determination of the direction of
upgoing tracks with resolutions better than 0.5$^\circ$.

The main goal of the experiment is to search for high energy neutrinos
with energies greater than 100~GeV by detecting muons produced in
neutrino charged current interactiona taking place in the vicinity of
the detector. Due to the large background from downgoing atmospheric
muons, the telescope is optimized for the detection of upgoing muons
that can only originate from neutrinos.

\section{Search for point sources}

A search for point sources has been carried out using the data taken
by ANTARES from 2007 to 2010.  Data runs were selected requiring that
most of the detector was operating and that the optical background
from bioluminiscence was low in terms of the baseline rate and burst
fraction. The final data sample amounted to a total of 813 live days.
Only events with upgoing muons were kept for further analysis,
requiring in addition that the corresponding track had a good
reconstruction quality and an estimated angular error lower than
1$^{\circ}$. The cut in quality was chosen so as to optimize the
discovery potential. A total of 3058 events were selected. According
to Monte Carlo simulations around 15\% of them were atmospheric muons
wrongly reconstructed as upgoing tracks.

\begin{figure}[htb]
\begin{center}
\includegraphics[scale=0.50]{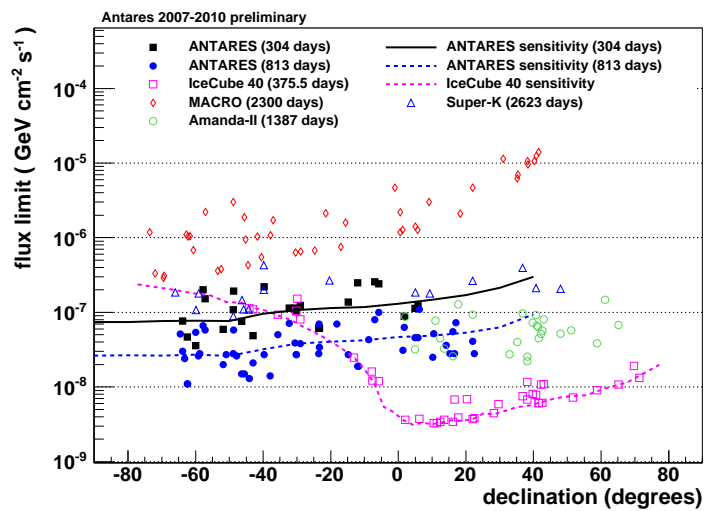}
\end{center}
\caption{\label{pslimits} 90\% CL upper limits for a neutrino flux
  with an E$^{-2}$ spectrum for 51 candidates sources (blue points)
  and the corresponding sensitivity (dashed blue curve). Results from
  the MACRO, Amanda~II, Super Kamiokande and IceCube
  telescopes~\cite{psothers} are also shown.  }
\end{figure}

Clusters of events with a significance above that expected from
background fluctuations were looked for with a likehood ratio method.
The likehood used the distribution in declination of the atmospheric
background obtained scrambling the data in right ascension and an
angular resolution of (0.46$\pm$0.15)$^{\circ}$, as given by Monte
Carlo simulation.  The full sky was searched for possible sources and
then a list of 51 pre-selected directions in the sky corresponding to
possible astrophysical neutrino sources were scrutinized. No
significant excess was found in either case.  An alternative search
method~\cite{em} was used as a cross-check obtaining similar results.
The post-trial $p$-value of the most significant cluster (at $\alpha =
-46.5^{\circ}$ and $\delta= -65.0^{\circ}$) was 2.5\% for the full sky
search --a value not significant enough to claim a signal--, whilst
the most significant source of the predefined list (HESS J1023-575)
was fully compatible with a background fluctuation ($p$=41\%). The
corresponding limits for neutrino sources emitting with an $E^{-2}$
energy spectrum are given in Figure~\ref{pslimits}, together with
limits from other experiments~\cite{psothers}. As can be seen, our
results are at present the more stringent for the Southern Sky, except
for the case of the IceCube detector for which in this hermisphere
very high energy neutrinos are looked for (E$>$ 1~PeV).  Our present
limit is 2.5 times better than our previous publish
result~\cite{psprevious}.

\section{Multimessenger searches}

The search of neutrinos in coincidence with other messengers has
several advantages. Sources already known to have high-energy
emission, e.g. gamma-rays, can be investigated, increasing the chance
to observe sites of hadronic acceleration. In addition, the
restriction of the search to limited time windows and sky directions
highly reduces the atmospheric neutrino background and therefore
increases the sensitivity to possible signals, so that a handful of
events can be enough to claim a signal. In the case of neutrino events
coincident with gravitational waves, the same astrophysical phenomena
are expected to produce both types of signals.  We give below a couple
of examples of this multimessenger program in ANTARES, which is too
broad to be fully reported here.

A selection of flares from blazars observed by the LAT detector of the
Fermi satellite during 2008 was carried out and the data taken by
ANTARES in the same period investigated for neutrino coincidences with
the flaring period of the blazars~\cite{blazars}. The selected blazars
are shown in Table~\ref{tab:blazars} together with the number of
events required to claim a 5$\sigma$ signal. Only one event --during a
flare of 3C279-- was detected. The post-trial p-value of such a
coincidence is 10\%, that is compatible with a background
fluctuation. The 90\% CL limits on the neutrino fluence from these
blazars are given in Table~\ref{tab:blazars}.

\begin{table}[ht!]
      \begin{center}
      \begin{tabular}{|c|c|c||c|c|c|}
      \hline 
Source & {N($5\sigma$)} & {Fluence} & Source &
             {N($5\sigma$)} & {Fluence}\\ \hline \hline {PKS0208-512}
             & 4.5 & 2.8 & {AO0235+164} & 4.3 & 18.7\\ \hline
             {PKS1510-089} & 3.8 & 2.8 & {3C273} & 2.5 & 1.1 \\ \hline
             {3C279} & 5.0 & 8.2 & {3C454.3} & 4.4 & 23.5 \\ \hline
             {OJ287} & 3.9 & 3.4 & {PKS0454-234} & 3.3 & 2.9 \\ \hline
             {WComae} & 3.8 & 3.6 & {PKS2155-304} & 3.7 & 1.6
             \\ \hline
      \end{tabular}
\caption{List of blazars for which neutrinos were looked for in
  coincidence with their flares. {\it N(5$\sigma$)} is the average
  number of events required for a 5$\sigma$ discovery (50\%
  probability) and {\it Fluence} is the upper limit (90\% CL) on the
  neutrino fluence in GeV$\cdot$cm$^{-2}$.}
      \label{tab:blazars}
      \end{center}     
\end{table}
\noindent
Several models predict the production of high energy neutrinos during
gamma-ray bursts. As in the previous analysis, restricting the search
to a short time window sizeably reduces the atmospheric background so
that only a few events would be enough to claim a discovery. Using the
2007 ANTARES data, a search for neutrinos coming from 37 GRBs events
was performed. No neutrino event was found in the corresponding time
windows and within the defined search cone around each source. An
overall limit of E$^{-2} \Phi_{\nu} <$ 1.8 $\times$ 10$^{-3}$ GeV
cm$^{-2}$ s$^{-1}$ has been set.

\section{Search for exotic particles}

ANTARES could detect a variety of hypothetical particles thanks to its
sensitivity to light emission. Under certain conditions monopoles
would leave an extremely conspicuous signal in ANTARES.  The existence
of monopoles has been put forward in the context of several
theories. Presently, there is no clear evidence of their existence and
several limits have been set on the flux of monopoles crossing the
Earth. Magnetic charges crossing the sea water at a speed larger than
their Cherenkov threshold ($\beta > 0.74$ in water) would produce a
huge amount of light. For one unit of magnetic charge this radiation
would be 8550 times larger than that of a muon. Moreover, even below
the threshold the ionization electrons produced by the monopole would
also radiate a large amount of light.

\begin{figure}[htb]
\begin{center}
\includegraphics[scale=0.55]{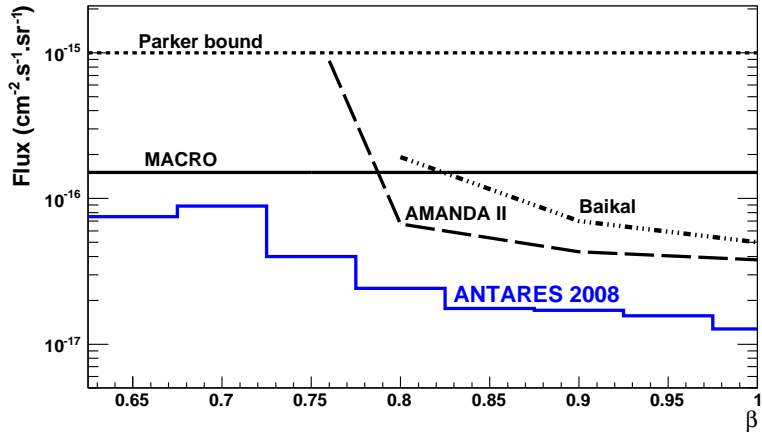}
\end{center}
\caption{\label{monopoles} 90\% CL upper limit on a flux of magnetic
  monopoles with velocities 0.625 $\le \beta \le$ 0.995.  Previous
  limits from the Amanda~II, Baikal and MACRO
  experiments~\cite{monopolesothers} are also shown. }
\end{figure}
\noindent

Using the data taken by ANTARES during 2007 and 2008, a search for
magnetic monopoles has been performed based on the number of hits and
the reconstructed velocity of the tracks. The selection criteria were
optimized for discovery in eight velocity intervals in the region
0.625 $\le \beta \le$ 0.995. Only one candidate was found, compatible
with the total expected background. In Figure~\ref{monopoles} the 90\%
CL upper limit on the flux of upgoing monopoles obtained is
shown~\cite{monopoles}. As can be seen, this limit is more stringent
than the previous existing limits~\cite{monopolesothers}.

A search for nuclearites, massive aggregates of up, down and strange
quarks, has also been performed. Nuclearites would produce in water a
thermal shock wave emitting a large amount of radiation at visible
wavelengths. No clear indication of nuclearites was observed using the
2007-2008 data sample and a 90\% CL upper limit of 10$^{-16}$
cm$^{-2}$ sr$^{-1}$ s$^{-1}$ for a flux of nuclearites with masses
between 10$^{-14}$ and 10$^{-17}$ GeV was establish.

\section{Summary}
Although no signal of cosmic neutrinos has been detected yet, the
results of the search for point sources indicates that the ANTARES
telescope has reached the expected performances, in particular an
angular resolution better than 0.5$^{\circ}$. Its sensitivity is at
present the best for neutrino sources in the 100~GeV to 100~TeV energy
range in the Southern Sky. The limits reported here on neutrino point
sources in the Southern Sky and on monopoles and nuclearites are the
most stringent up to date.

\section*{Acknowledgements}
We gratefully acknowledge the financial support of the Spanish
Ministerio de Ciencia e Innovaci\'on (MICINN), grants
FPA2009-13983-C02-01, ACI2009-1020 and Consolider MultiDark
CSD2009-00064 and of the Generalitat Valenciana, Prometeo/2009/026.

\end{document}